\documentclass[pra,superscriptaddress,12pt]{revtex4}
\usepackage{hyperref,graphicx}
\usepackage{amsthm,amssymb,amsmath,bbm}

\begin{document}
\title{A note on one-way quantum deficit and quantum discord}

\author{Biao-Liang Ye}
\affiliation{School of Mathematical Sciences, Capital Normal University,
Beijing 100048, China}
\author{Shao-Ming Fei}
\affiliation{School of Mathematical Sciences, Capital Normal University,
Beijing 100048, China}
\affiliation{Max-Planck-Institute for Mathematics in the Sciences,
Leipzig 04103, Germany}

\begin{abstract}
  {\bf Abstract} One-way quantum deficit and quantum discord are two important
  measures of quantum correlations.
  We revisit the relationship between them in two-qubit systems. We
  investigate the conditions that both one-way quantum deficit and quantum discord
  have the same optimal measurement ensembles, and demonstrate that one-way quantum deficit
  can be derived from the quantum discord for a class of $X$ states.
  Moreover,
  we give an explicit relation between one-way quantum deficit and
  entanglement of formation. We show that under
  phase damping channel both one-way quantum deficit and quantum discord evolve
  exactly in the same way for four parameter $X$ states.
  Some examples are presented in details.

\end{abstract}
\maketitle

\noindent{\bf 1 Introduction}

Quantum entanglement plays important roles in quantum information
and quantum computation \cite{Horodecki2009}.
However, some quantum states without quantum entanglement
can also perform quantum tasks \cite{Datta2008,Lanyon2008} like
quantum state discrimination \cite{Roa2011,Li2012}, remote state preparation \cite{dakic2012quantum},
quantum state merging \cite{Madhok2011,Cavalcanti2011} etc., which have led to new definitions of quantum correlations such as
quantum discord \cite{Ollivier2001,Henderson2001}, one-way quantum deficit \cite{Oppenheim2002,Horodecki2003,Devetak2005,Horodecki2005}, and various
`discord like' measures \cite{modi2012}.

One-way quantum deficit was
first proposed by Oppenheim et al \cite{Oppenheim2002} for studying thermodynamical
systems.
They considered the amount of work which could be extracted from
a heat bath by local operations.
It quantifies the minimum distillable entanglement generated
between the whole system and the measurement apparatus in measuring
one subsystem of the whole system \cite{Streltsov2011}.
The analytical formulae of one-way quantum deficit is not known even
for two-qubit states. With limited analytical results \cite{Wang2013,Wang2015}, many discussions on quantum
deficit only rely on numerical results,
since it involves minimization of sum of local and conditional entropies.

Another famous measure of quantum correlations,
the quantum discord \cite{Ollivier2001,Henderson2001}, is defined to be the difference of two classically
equivalent expressions for the mutual information.
There have been a lot of results on quantum discord for bipartite as well as
multipartite mixed quantum states \cite{modi2012}. Nevertheless, due to the optimization problem involved,
it has been recently shown that calculating quantum discord is
an NP complexity problem \cite{Huang2014}.

It is meaningful to link directly one-way quantum deficit
to quantum discord. The relationship between quantum discord and one-way
quantum deficit was first discussed in Ref. \cite{Horodecki2005}.
Horodecki {\it et al}  shew that the one-way quantum deficit is upper bounded by
the quantum discord for any bipartite quantum states.
In Ref. \cite{zhu2013one} a tradeoff relationship
between one-way unlocalizable quantum discord and
one-way unlocalizable quantum deficit has been presented.
The tradeoff relationship between quantum discord and one-way quantum
deficit is obtained \cite{xiquantum2015}.

Anyway, decisive results between quantum discord
and one-way quantum deficit is not fully explored even for the
two-qubit $X$ states yet.
Here, we revisit the relationship between
one-way quantum deficit and quantum
discord. We find that for special two-qubit $X$ states the one-way quantum deficit can be derived
from quantum discord exactly in some optimal measurement bases. Furthermore,
we connect one-way quantum deficit to
entanglement of formation directly.

To capture the non-classical correlations in bipartite systems,
let us recall the following two popular measures of quantum correlations.

\emph{One-way quantum deficit}
Suppose Alice and Bob are allowed to
perform only local operations. Consider a one-way classical
communication, say, from Alice to Bob.
The amount of information extractable from quantum system $\varrho^{AB}$ is given
by $\mathcal{I}_e=\log_2 \mathcal{D}-S(\varrho^{AB})$,
where $\mathcal{D}$ is the dimension of the Hilbert space,
$S(\varrho)=-{\rm Tr}[\varrho \log_2 \varrho]$ is the
von Neumann entropy of a quantum state $\varrho$.

The classical operations to extract the amount of information from the quantum state is
$\mathcal{I}_o=\log_2 \mathcal{D}-\min S((\varrho^{AB})')$, where
$(\varrho^{AB})'=\sum_k M_k^A\varrho^{AB} M_k^A$ is the
quantum state after measurement $M_k^{A}$ has been
performed on $A$. The one-way quantum deficit \cite{Oppenheim2002,Horodecki2003,Devetak2005,Horodecki2005} is given by
the difference  of $\mathcal{I}_e$ and $\mathcal{I}_o$ \cite{Streltsov2011},
\begin{eqnarray}
\overset{\rightharpoonup}{\vphantom{a}\smash{\Delta}}
&=&\mathcal{I}_e-\mathcal{I}_o \nonumber\\
&=&\min S(\sum_k M_k^A\varrho^{AB} M_k^A)-S(\varrho^{AB}).
\end{eqnarray}
The minimum is taken over all local measurements $M_k^{A}$.
This quantity is equal to the thermal discord \cite{Zurek2003}.

\emph{Quantum discord} The quantum discord
is defined as the minimal difference between quantum mutual
information and classical correlation. The quantum mutual information
is denoted by $\mathcal{I}(\varrho^{AB})=S(\varrho^A)+S(\varrho^B)-S(\varrho^{AB})$,
which is also identified as the total correlation of the bipartite
quantum system $\varrho^{AB}$. The $\varrho^{A(B)}$
are the reduced density matrices ${\rm Tr}_{B(A)}\varrho^{AB}$, respectively.
Let $\{M_k^A\}$ be a measurement on subsystem $A$.
Classical correlation is given as $\mathcal{J}(\varrho^{AB})=S(\varrho^B)-\min \sum_k p_kS(\varrho_{M_k^A}^{B})$,
where $p_k={\rm Tr}(M_k^A\otimes I_2 \varrho^{AB})$ is the probability of $k$th measurement outcome 
and $\varrho_{M_k^A}^{B}={\rm Tr}_A[M_k^A\otimes I_2 \varrho^{AB}]/p_k$
is the post-measurement state.

The quantum discord \cite{Ollivier2001,Henderson2001} is defined by
\begin{eqnarray}
\overset{\rightharpoonup}{\vphantom{a}{\smash{\delta}}}
&=&\mathcal{I}(\varrho^{AB})-\mathcal{J}(\varrho^{AB})\nonumber\\
&=&S(\varrho^A)+\min \sum_k p_k S(\varrho_{M_k^A}^{B})-S(\varrho^{AB}).
\end{eqnarray}
The superscript $``\rightharpoonup"$ stands for that the measurement performed on subsystem $A$.
The minimum is taken over all possible measurements $\{M_k^A\}$ on the subsystem $A$.\\


\noindent{\bf 2 Linking one-way quantum deficit to quantum discord}

Let us consider bipartite systems in Hilbert space $\mathbbm{C}^2\otimes \mathbbm{C}^2$.
Generally, the quantum correlations
are invariant under local unitary operations \cite{modi2012}.
Hence, one can write the $X$ states \cite{Yurischev2015} in the form
\begin{eqnarray}\label{3}
\varrho^{AB}=\frac14(I_2\otimes I_2+a\sigma_z\otimes I_2+bI_2\otimes\sigma_z
+\sum_{i\in\{x,y,z\}}c_i\sigma_i\otimes\sigma_i),
\end{eqnarray}
where $\sigma_i (i\in\{x, y, z\})$ are Pauli matrices,
$I_2$ is the identity matrix, and the parameters
$\{a, b, c_x, c_y,c_z\}\in[-1, 1]$ are real numbers.

The optimal measurement with measurement operators satisfying
$M_k^A\geqslant0$, $\sum_kM_k^A=I$, are generally positive operator-valued measurement (POVM). For rank-two
two-qubit systems, the optimal measurement is just
projective ones \cite{Shi2012}. It is also sufficient to consider
projective measurement for rank-three and four \cite{Galve2011}.

Let $M_k^A=|k'\rangle\langle k'|$,
$k\in\{0, 1\}$,  where
\begin{eqnarray}
&&|0'\rangle=\cos(\theta/2)|0\rangle-e^{-i\phi}\sin(\theta/2)|1\rangle,\\
&&|1'\rangle=e^{i\phi}\sin(\theta/2)|0\rangle+\cos(\theta/2)|1\rangle.
\end{eqnarray}

For the given system (\ref{3}), we obtain
$\overset{\rightharpoonup}{\vphantom{a}\smash{\delta}}=S(\varrho^{A})+\min \sum_k p_k S(\varrho_{M_k^A}^{B})-S(\varrho^{AB})$,
in which
\begin{eqnarray}
p_{k\in\{0,1\}}=\frac12(1\pm  a\cos \theta),
\end{eqnarray}
$S(\varrho^{A})=h(\frac{1+a}{2})$ with $h(x)=-x\log_2x-(1-x)\log_2(1-x)$,
\begin{eqnarray}\label{condis}
\sum_k p_kS(\varrho_{M_k^A}^{B})&&=p_0S(\varrho_{M_0^A}^{B})+p_1S(\varrho_{M_1^A}^{B})\nonumber\\
  &&=-\sum\limits_{k,j\in\{0,1\}}p_k w_{kj} \log w_{kj},
\end{eqnarray}
with $w_{00}, w_{01}$ and $w_{10}, w_{11}$ the eigenvalues of $\varrho_{M_0^A}^{B}$ and
$\varrho_{M_1^A}^{B}$, respectively,
\begin{equation}
w_{kj\in\{0,1\}}=\{1+(-1)^ka \cos \theta +(-1)^j\sqrt{[c_x^2\cos ^2\phi +c_y^2\sin ^2\phi ] \sin ^2\theta  +[b+(-1)^kc_z \cos \theta ]{}^2}\}/(4p_k).
\end{equation}

The corresponding quantity $S(\sum_k M_k^A\varrho^{AB} M_k^A)$ in the definition of one-way quantum deficit
is given by
\begin{eqnarray}\label{minterm}
S(\sum_k M_k^A\varrho^{AB} M_k^A)
  &&=S(M_0^A\otimes p_0\varrho_{M_0^A}^{B}+M_1^A\otimes p_1\varrho_{M_1^A}^{B})\nonumber\\
  &&=S(p_0\varrho_{M_0^A}^{B})+S(p_1\varrho_{M_1^A}^{B})\nonumber\\
  &&=-\sum\limits_{k,j\in\{0,1\}}p_k w_{kj} \log p_kw_{kj}\nonumber\\
  &&=h(p_0)-\sum\limits_{k,j\in\{0,1\}}p_k w_{kj} \log w_{kj}.
\end{eqnarray}
Substituting Eq. (\ref{condis}) into above equation, we have
\begin{eqnarray}
S(\sum_k M_k^A\varrho^{AB} M_k^A)=h(p_0)+\sum_k p_kS(\varrho_{M_k^A}^{B}),
\end{eqnarray}
which is joint entropy theorem \cite{Shao2013}.

Let us set
\begin{eqnarray}
\mathcal{F}&=&S(\varrho^{A})+\sum_k p_k S(\varrho_{M_k^A}^{B})-S(\varrho^{AB}),\\
\mathcal{G}&=&S(\sum_k M_k^A\varrho^{AB}M_k^A)-S(\varrho^{AB}).
\end{eqnarray}
Inserting Eq.(\ref{condis}), Eq.(\ref{minterm}) into Eq.(11), Eq.(12), respectively,
we have
\begin{eqnarray}
\mathcal{F}
  &=&S(\varrho^{A})-\sum\limits_{k,j\in\{0,1\}}p_k w_{kj} \log w_{kj}-S(\varrho^{AB}),\\
\mathcal{G}
  &=&h(p_0)-\sum\limits_{k,j\in\{0,1\}}p_k w_{kj} \log w_{kj}-S(\varrho^{AB}).
\end{eqnarray}

To search for the minimization involved in computing quantum discord and
one-way quantum deficit is equivalent to seek for the minimal value of the function $\mathcal{F}$
and $\mathcal{G}$ with respect to the two
parameters $\theta$ and $\phi$ in the measurement operators.
Similar to the technique to minimize $\mathcal{F}(\theta, \phi)$, it is enough to consider minimize $\mathcal{F}(\theta, 0)$ in calculating the quantum discord of two-qubit $X$-states \cite{Maldonado-Trapp2015}. We denote
$$
G(\theta, \phi)=S(\sum_k M_k^A\varrho^{AB} M_k^A)=-\sum\limits_{k,j\in\{0,1\}}p_k w_{kj} \log p_kw_{kj}=-\sum_{l=1}^4\lambda_l\log_2\lambda_l,
$$
where,
\begin{eqnarray*}
\lambda_{1,2}=\frac{1}{4} \left(p_{0}\pm\sqrt{R+T_0}\right),~~~~
\lambda_{3,4}=\frac{1}{4} \left(p_{1}\pm\sqrt{R+T_1}\right),
\end{eqnarray*}
and $p_{0}=1+ a \cos \theta$, $p_{1}=1-a \cos \theta$,
$R=[c_x^2 \cos ^2\phi +c_y^2 \sin ^2\phi ]\sin ^2\theta$,
$T_{0}=\left(b+ c_z \cos \theta \right){}^2$, $T_{1}=\left(b-c_z \cos \theta \right){}^2$.
Since $\lambda_l\geqslant0$, one has $p_{k}\geqslant\sqrt{R+T_{k}}\geqslant0$.

Noting that $G(\theta, \phi)=G(\pi-\theta, \phi)=G(\theta, 2\pi-\phi)$ and $G(\theta, \phi)$ is symmetric
with respect to $\theta=\pi/2$ and $\phi=\pi$,
we only need to consider the case of $\theta\in[0, \pi/2]$ and $\phi\in[0,\pi)$.
The extreme points of $G(\theta, \phi)$ are determined by
the first partial derivatives of $G$ with respect to $\theta$ and $\phi$,
\begin{eqnarray}\label{cth}
\frac{\partial G}{\partial \theta}=-\frac{\sin\theta}{4}H_{\theta},
\end{eqnarray}
with
\begin{eqnarray}\label{optheta}
H_{\theta}=&&\frac{R\csc\theta\cot\theta-c_z
\sqrt{T_0}}{\sqrt{R+T_0}}\log_2\frac{p_0 +\sqrt{R+T_0}}{p_0 -\sqrt{R+T_0}}+a\log_2\frac{p_1^2-(R+T_1)}{p_0^2-(R+T_0)}\nonumber\\[2mm]
&&+\frac{R\csc\theta\cot\theta+c_z
\sqrt{T_1}}{\sqrt{R+T_1 }}\log_2\frac{p_1 +\sqrt{R+T_1}}{p_1 -\sqrt{R+T_1}},
\end{eqnarray}
and
\begin{eqnarray}\label{cph}
\frac{\partial G}{\partial \phi}=2\,ef\sin^2\theta\,\sin2\phi\, H_{\phi},
\end{eqnarray}
with
\begin{eqnarray}
H_{\phi}=\frac1{\sqrt{R+T_0}}\log_2\frac{p_0 +\sqrt{R+T_0}}{p_0 -\sqrt{R+T_0}}
+\frac1{\sqrt{R+T_1}}\log_2\frac{p_1 +\sqrt{R+T_1}}{p_1 -\sqrt{R+T_1}},
\end{eqnarray}
$e=\frac14|c_x+c_y|$ and $f=\frac14|c_x-c_y|$ where the absolute values have been taken
since the phase for $X$ states can be always
removed by local unitary operation \cite{modi2012}.

As $H_{\phi}$ is always positive, $\frac{\partial G}{\partial \phi}=0$ implies that
either $\phi=0, \pi/2$ for any $\theta$,
or $\theta=0$ for any $\phi$ which implies that Eq. (\ref{cth}) is zero and
the minimization is independent on $\phi$.
If $\theta\ne 0$, one gets the second derivative of $G$,
$$
\left.\frac{\partial^2G}{\partial\phi^2}\right|_{(\theta,0)}=4ef\sin ^2(\theta )H_{\phi=0}>0,
$$
and
$$
\left.\frac{\partial^2G}{\partial\phi^2}\right|_{(\theta,\pi/2)}=-4ef\sin ^2(\theta )H_{\phi=\pi/2}<0.
$$
Since for any $\theta$ the second derivative $\partial^2 G/\partial\phi^2$
is always negative for $\phi=\pi/2$, we only need to deal with the minimization problem
for the case of $\phi=0$. To minimize $G(\theta, \phi)$ becomes to minimize $G(\theta, 0)$.
Thus, we need only to find the minimal value of $\mathcal{F}$ and $\mathcal{G}$ by varying $\theta$ only.

Denote $\mathcal{F}(\theta)=\mathcal{F}|_{\phi=0}$, $\mathcal{G}(\theta)=\mathcal{G}|_{\phi=0}$ and
$\mathcal{H}(\theta)=\mathcal{G}(\theta)-\mathcal{F}(\theta)$.
The first derivative of $\mathcal{H}(\theta)$ with respect to $\theta$ is given by
\begin{eqnarray}
\mathcal{H}(\theta)'
&=&\frac{a}{2} \sin\theta \log_2 \frac{1+a \cos\theta}
{1-a \cos\theta}.
\end{eqnarray}
From $\mathcal{H}(\theta)'=0$, we have either $a=0$ or $\theta=0,\pi/2$. Since these stationary points make
$\mathcal{F}(\theta)'=\mathcal{G}(\theta)'$, they are the sufficient conditions
that both $\mathcal{G}(\theta)$ and $\mathcal{F}(\theta)$ reach the minimum with the same optimal measurement ensemble.
Here $a$ is a parameter of the $X$ states and $\theta$ is a parameter related to measurement.
Substituting $a=0$ or $\theta=0,\pi/2$ into $\mathcal{F}(\theta)$ and $\mathcal{G}(\theta)$ we have the following results:

\emph{Theorem} For two-qubit $X$ states, if the measurement is
performed on the subsystem $A$ (resp. $B$), then
$\overset{\rightharpoonup}{\vphantom{a}{\smash{\Delta}}}=\overset{\rightharpoonup}{\vphantom{a}{\smash{\delta}}}$
for $a=0$ (resp. $b=0$).
If their optimal measurement bases which depend on the parameters of the state are at $\theta=0$,
then $\overset{\rightharpoonup}{\vphantom{a}{\smash{\Delta}}}=\overset{\rightharpoonup}{\vphantom{a}{\smash{\delta}}}$.
If their optimal measurement bases which depend on the parameters of the state are at $\theta=\pi/2$, then
$\overset{\rightharpoonup}{\vphantom{a}{\smash{\Delta}}}=
\overset{\rightharpoonup}{\vphantom{a}{\smash{\delta}}}-S(\varrho^A)+1$.

Recently, we notice that in Ref \cite{Chanda2015}, the authors assumed that the quantum discord
and one-way quantum deficit get their minimal values in the
same measurement ensemble simultaneously. Thus similar to the quantum discord, the frozen quantum phenomenon
under bit flip channels of one-way quantum deficit happens.
Here, our \emph{Theorem} gives the explicit conditions that both quantum
discord and one-way quantum deficit have the same optimal measurement bases.

\emph{Corollary 1} The one-way quantum deficit is bounded by the quantum discord for two-qubit $X$ states,
\begin{equation}\label{19p}
\overset{\rightharpoonup}{\vphantom{a}\smash{\delta}}\leqslant\overset{\rightharpoonup}{\vphantom{a}\smash{\Delta}}
\leqslant S(\rho^A).
\end{equation}

\begin{proof}
Since $0\leqslant\mathcal{H}\leqslant1$, we have
$\overset{\rightharpoonup}{\vphantom{a}\smash{\delta}}\leqslant\overset{\rightharpoonup}{\vphantom{a}\smash{\Delta}}
\leqslant \overset{\rightharpoonup}{\vphantom{a}\smash{\delta}}+1$ . By using the tight bound about one-way quantum deficit
$\Delta\leqslant S(\rho^A)$ in Ref. \cite{Shao2013}, we obtain (\ref{19p}).
\end{proof}

\emph{Corollary 2} One-way quantum deficit and the entanglement of formation satisfy the following relations for two-qubit $X$ states,
\begin{eqnarray}\label{co}
\overset{\rightharpoonup}{\vphantom{a}{\smash{\Delta}}}
&&=E_f(\varrho^{BC})-S(\varrho^{AB})
+\left\{
   \begin{array}{ll}
     1, & \hbox{a=0 or $\theta=\pi/2$;} \\
     h(\frac{1-a}{2}), & \hbox{$\theta=0$,}
   \end{array}
 \right.
\end{eqnarray}
where $C$ is the assisted system to purify the state $\varrho^{AB}$,
and $E_f(\varrho^{BC})$ is the entanglement of formation of
$\varrho^{BC}$, while $\varrho^{BC}$ is the reduced state
from a pure state $|\psi\rangle_{ABC}$.

\begin{proof}
From the Koashi-Winter equality \cite{Koashi2004}
\begin{equation}
S(\varrho^{B})=\mathcal{J}(\varrho^{AB})+E_f(\varrho^{BC}),
\end{equation}
and $\mathcal{J}(\varrho^{AB})=S(\varrho^B)-\min \sum_k p_kS(\varrho_{M_k^A}^{B})$,
one has $E_f(\varrho^{BC})=\min \sum_k p_kS(\varrho_{M_k^A}^{B})$.
Consequently, quantum discord is rewritten as
\begin{equation}\label{ef}
\overset{\rightharpoonup}{\vphantom{a}{\smash{\delta}}}=S(\varrho^{A})+E_f(\varrho^{BC})-S(\varrho^{AB}).
\end{equation}
Thus, we have
\begin{eqnarray}
\overset{\rightharpoonup}{\vphantom{a}{\smash{\Delta}}}=
\overset{\rightharpoonup}{\vphantom{a}{\smash{\delta}}}
=h(\frac{1-a}{2})+E_f(\varrho^{BC})-S(\varrho^{AB}),
\end{eqnarray}
where both of the optimal measurement bases are taken at $\theta=0$.
Hence for $a=0$, we have $\overset{\rightharpoonup}{\vphantom{a}{\smash{\Delta}}}=E_f(\varrho^{BC})-S(\varrho^{AB})+1$ indeed.
For $\theta=\pi/2$, by using the relations in \emph{Theorem} and Eq.(\ref{ef}) we also get
$\overset{\rightharpoonup}{\vphantom{a}{\smash{\Delta}}}=E_f(\varrho^{BC})-S(\varrho^{AB})+1$.
\end{proof}

\emph{Remark} Recently, in Ref. \!\!\cite{xiquantum2015} by using measure of relative entropy of coherence,
\begin{equation}
C_{RE}(\varrho^A)=\min_{\sigma \in \mathcal{I}} S(\varrho^A||\sigma),
\end{equation}
where $\mathcal{I}$ stands for the set of decoherence states $\sigma=\sum_i\mu_i|i\rangle\langle i|$ with $\mu_i\in[0,1]$ and $\sum_i\mu_i=1$,
the authors provided a tradeoff relationship between
$\overset{\rightharpoonup}{\vphantom{a}{\smash{\delta}}}$ and $\overset{\rightharpoonup}{\vphantom{a}\smash{\Delta}}$, i. e.,
$\overset{\rightharpoonup}{\vphantom{a}{\smash{\delta}}}+C_{RE}(\varrho^{A})
=\overset{\rightharpoonup}{\vphantom{a}\smash{\Delta}}$.

In fact, one-way quantum deficit can be derived from
quantum discord directly.
We consider the exact relationship
between quantum discord and one-way quantum deficit
in the following examples.

\emph{Example 1}.
The Bell-diagonal state
$\varrho_{Bell}^{AB}=\frac{1}{4}(I_2\otimes I_2+\sum_{i\in\{x,y,z\}}c_i\sigma_i\otimes\sigma_i)$. In this case
$a=0$ and
\begin{eqnarray}
\overset{\rightharpoonup}{\vphantom{a}{\smash{\Delta}}}
=\overset{\rightharpoonup}{\vphantom{a}{\smash{\delta}}}
=h(\frac{1-c}{2})+\sum_{s\in\{jkl\}}A_s\log_2 A_s,
\end{eqnarray}
where
$s$ is the set $ \{jkl\}=\{111,100,010,001\}$,
$A_{jkl}=\frac14(1+(-1)^jc_x+(-1)^kc_y+(-1)^lc_z)$,
and $c\equiv\max\{|c_x|, |c_y|, |c_z|\}$.
Therefore, from \emph{Theorem} we get the analytical expression of one-way
quantum deficit from quantum discord given in \cite{Luo2008a}.

{\it Example 2}. Consider a class of X-state,
\begin{equation}
  \varrho_q^{AB}=q|\psi^-\rangle\langle\psi^-|+(1-q)|00\rangle\langle00|,
\end{equation}
where $|\psi^-\rangle=\frac1{\sqrt{2}}(|01\rangle-|10\rangle$).

For this state, quantum discord is derived
at $\theta=\pi/2$ for $q\in[0,1]$. The optimal basis of one-way
quantum deficit for $q\in[0.67,1]$ is also at $\theta=\pi/2$. The
value 0.67 is the solution of $H_{\theta}'|_{\theta=\pi/2,\phi=0}=0$ in Eq. (\ref{optheta})
for the state $\varrho_q^{AB}$. According to the \emph{Theorem},
we have
\begin{eqnarray}
\overset{\rightharpoonup}{\vphantom{a}{\smash{\Delta}}}=E_f(\varrho^{BC})-S(\varrho^{AB})+1,
\end{eqnarray}
where the entanglement of formation
\begin{eqnarray}
E_f(\varrho^{BC})=h(\frac{1+\sqrt{1-\mathcal{C}^2}}{2})
\end{eqnarray}
with concurrence $\mathcal{C}=\sqrt{2q(1-q)}$.
So analytical one-way quantum deficit of state $\varrho_q^{AB}$ is
\begin{eqnarray}
\overset{\rightharpoonup}{\vphantom{a}{\smash{\Delta}}}
=h(\frac{1+\sqrt{1-\mathcal{C}^2}}{2})-h(q)+1
\end{eqnarray}
for $q\in[0.67, 1]$, see Fig. 1.

\begin{figure}[h]
\includegraphics[width=8.1cm]{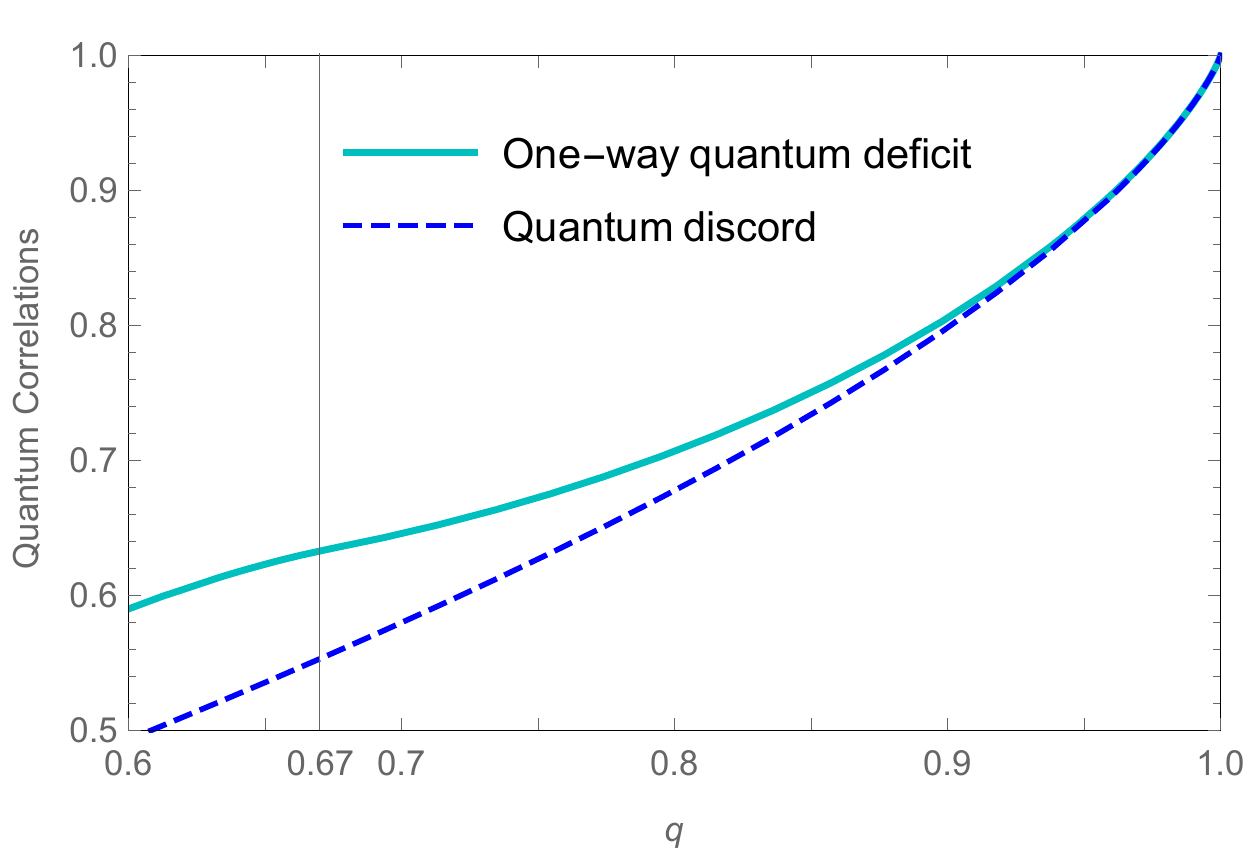}\\
\centering
\caption{One-way quantum deficit (turquoise solid line) and quantum discord (blue dashed line) vs $q$.
The interval $q\in[0.67, 1]$ one-way quantum deficit and quantum discord both get their optimum at $\theta=\pi/2$.}
\label{QQ}
\end{figure}

\noindent{\bf 3 Quantum correlations under phase damping channel}

A quantum system would be subject to interaction with environments.
We consider now the evolution of one-way quantum deficit and quantum discord
under noisy channels.
Consider a class of initial two-qubit states,
\begin{equation}
\Omega=\frac14(I_2\otimes I_2+bI_2\otimes\sigma_z
+\sum_{i\in\{x,y,z\}}c_i\sigma_i\otimes\sigma_i).
\end{equation}

If both two qubits independently goes through a channel
given by the Kraus operators $\{K_i\}$, $\sum_i K_i^\dagger K_i=I$.
The state $\Omega$ evolves into
\begin{equation}
\tilde{\Omega}=\sum_{i,j\in\{1,2\}}K_i^A\otimes K_j^B\cdot \Omega \cdot [K_i^A\otimes K_j^B]^\dagger.
\end{equation}
For phase damping channels \cite{nielsen2010quantum}, the Kraus operators are given by $K_1^{A(B)}=|0\rangle\langle0|+\sqrt{1-\gamma}|1\rangle\langle1|$, and
$K_2^{A(B)}=\sqrt{\gamma}|1\rangle\langle1|$ with the decoherence rate $\gamma\in[0,1]$.
Thus we have
\begin{equation}
\tilde{\Omega}=\frac14[I_2\otimes I_2+bI_2\otimes\sigma_z
+c_z\sigma_z\otimes\sigma_z+\sum_{i\in\{x,y\}}(1-\gamma)c_i\sigma_i\otimes\sigma_i],
\end{equation}
which is a two-qubit $X$ state with $a=0$. From the \emph{Theorem}, we obtain
one-way quantum deficit and quantum discord performed on the subsystem $A$
evolves coincidentally with each other all the time.

\begin{figure}[h]
  \includegraphics[width=8.1cm]{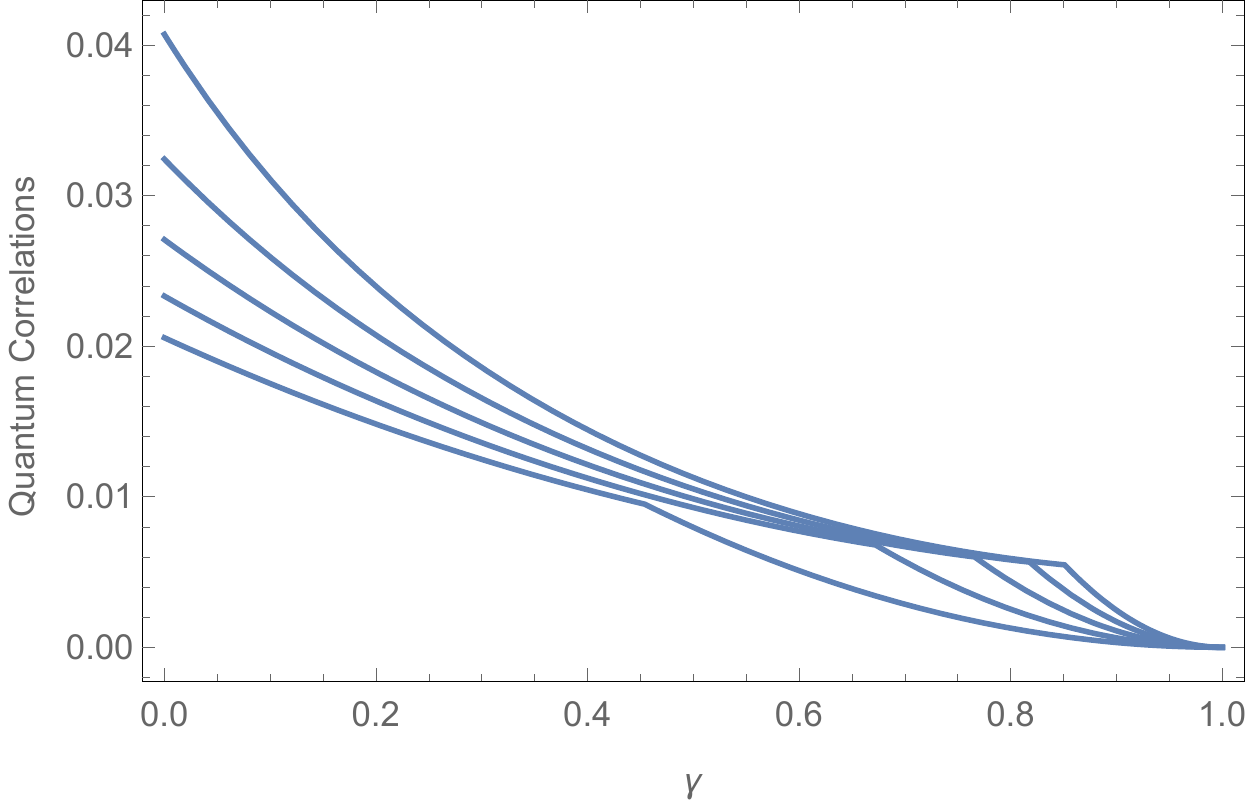}\\
  \caption{One-way quantum
  deficit and quantum discord evolve exactly in the same way under
  phase damping channel. The solid line from bottom to top correspond $c_y=0.15, 0.25, 0.35, 0.45, 0.55$ respectively, for fixed  parameters $b=0.26, c_x=0.13$, and $c_z=0.08$.}\label{phasenoise}
\end{figure}

For example, we draw the quantum discord and one-way quantum deficit vs parameter $\gamma$
in Fig. \ref{phasenoise} for $b=0.26$, $c_x=0.13$, $c_z=0.08$, and
$c_y=0.15, 0.25, 0.35, 0.45, 0.55$ respectively.

\noindent{\bf 4 Conclusions}

We have investigated the connections between one-way quantum deficit
and quantum discord for two-qubit $X$ states.
Sufficient conditions are given that the one-way quantum deficit can be
derived from quantum discord directly.
The explicit relation between one-way quantum deficit and entanglement
of formation is also presented. Moreover,
we have shown that the one-way quantum deficit
and quantum discord of a class of four parameters $X$ states evolve
coincidentally under phase damping channel.
Our results may enlighten the understanding
on the relations between one-way quantum deficit and quantum discord.
It is also interesting to study the relationship between one-way quantum deficit and quantum discord for higher dimensional
and multipartite systems.\\

\noindent {\bf Acknowledgements }
We thank the anonymous referees for their careful reading and valuable comments.
This work is supported by NSFC under number 11275131.\\


\end{document}